\documentclass[12pt]{article}

\usepackage{scicite}
\usepackage{graphicx}
\usepackage{times}
\usepackage{makecell}
\usepackage{comment}

\usepackage{amsmath}
\usepackage{cases}
\usepackage{ulem}
\usepackage{color}
\usepackage{xcolor,colortbl}
\definecolor{r}{rgb}{1,0,0}   
\definecolor{g}{rgb}{0,1,0}   
\definecolor{b}{rgb}{0,0,1}

\definecolor{variables}{rgb}{1.0,1.0,1.0}
\definecolor{softness}{rgb}{0.55,0.75,0.05}
\definecolor{kappa}{rgb}{0,0.67,0.86}
\definecolor{zeta}{rgb}{0.57,0.44,0.90}
\definecolor{itac}{rgb}{0.95,0.35,0.14}
\definecolor{cd2min}{rgb}{0.98,0.69,0.23}
\definecolor{highlight}{rgb}{0.95,0.95,0.95}

\newenvironment{sciabstract}{%
\begin{quote} \bf}
{\end{quote}}

\newcounter{lastnote}

\title{Machine learning-informed structuro-elastoplasticity predicts ductility of disordered solids} 

\author
{
Hongyi Xiao,$^{1,2,3\dagger}$ Ge Zhang,$^{1,4\dagger}$, Entao Yang,$^{5\dagger}$ Robert J. S.  Ivancic,$^{6\dagger}$\\ Sean A. Ridout,$^{1}$ Robert Riggleman,$^{5}$ Douglas J. Durian,$^{1}$ and Andrea J. Liu$^{1\ast}$\\
\\
\normalsize{$^{1}$Department of Physics and Astronomy, University of Pennsylvania, Philadelphia, PA, USA}\\
\normalsize{$^{2}$Institute for Multiscale Simulation, Friedrich-Alexander-Universit\"{a}t Erlangen-N\"{u}rnberg, } 
\\ \normalsize{Erlangen, Germany}
\\ \normalsize{$^{3}$Department of Mechanical Engineering, University of Michigan, Ann Arbor, MI, USA}\\
\normalsize{$^{4}$Department of Physics, City University of Hong Kong, Hong Kong, China}\\
\normalsize{$^{5}$Department of Chemical and Biomolecular Engineering, University of Pennsylvania,}
\\ \normalsize{Philadelphia, PA, USA}\\
\normalsize{$^{6}$Materials Science and Engineering Division, National Institute of Standards and Technology, }
\\ \normalsize{Gaithersburg, MD, USA}\\
\normalsize{$^\dagger$Equal contribution.}
\\
\normalsize{$^\ast$To whom correspondence should be addressed; E-mail:  ajliu@physics.upenn.edu.}
}

\date{}

\begin{document} 

\baselineskip24pt



\maketitle 




\begin{sciabstract}
All solids yield under sufficiently high mechanical loads. Below yield, the mechanical responses of all disordered solids are nearly alike, but above yield every different disordered solid responds in its own way. Brittle systems can shatter without warning, like ordinary window glass, or exhibit strain localization prior to fracture, like metallic or polymeric glasses. Ductile systems, e.g. foams like shaving cream or emulsions like mayonnaise, can flow indefinitely with no strain localization. 
While there are empirical strategies for tuning the degree of strain localization, there is no framework that explains their effectiveness or limitations. We show that Structuro-Elastoplastic (StEP) models provide microscopic understanding of how strain localization depends on the interplay of structure, plasticity and elasticity.

\end{sciabstract}



Disordered solids, such as metallic, molecular, polymeric, nanoparticle or colloidal glasses or granular packings, exhibit plastic behaviors distinct from those of crystalline solids~\cite{falk2011deformation, Greer2013, bonn2017yield, tanguy2021elasto}. 
Below yield, these behaviors are surprisingly universal with a consistent value of yield strain~\cite{cubuk2017structure}. Beyond yield, however, some disordered solids, such as foams and certain polymer glasses, are ductile with little or no strain localization, while others, like molecular and metallic glasses, typically exhibit sharp strain localization in shear bands as precursors to brittle failure. 

Several empirical strategies have been discovered for tuning strain localization. Reducing the range of inter-particle attractions~\cite{falk1999molecular, dauchot2011athermal, lin2019distinguishing, xiao2020strain, karmakar2011effect}, equilibrating better~\cite{ozawa2018random}, annealing at slower cooling rates~\cite{shavit2014strain} or cooling while loading~\cite{matsushige1976pressure,lin2019distinguishing} all enhance strain localization. Friction~\cite{karimi2019plastic}, composite constituents~\cite{wang2019stretchable}, particle shape~\cite{zhang2013using} and degree of crystallinity~\cite{dauchot2011athermal} also influence ductility. However, we do not understand why or how these factors influence strain localization. 


At the microscopic scale, plasticity in solids is accomplished by rearrangements in which constituent particles change neighbors. 
While other approaches exist~\cite{sollich1997rheology,barlow2020ductile,falk1998dynamics,hinkle2017coarse}, elastoplasticity (EP) models~\cite{nicolas2018deformation} update and record yield strain and strain at each lattice site, corresponding to a coarse-grained region. Such models typically assume an underlying distribution of local yield strains that controls the degree of strain localization~\cite{popovic2018elastoplastic} and is put in by hand~\cite{nicolas2018deformation,tanguy2021elasto}.  


Here we follow a different path. 
A particle's local yield strain--and hence its probability to rearrange--depends on its local structural environment~\cite{barbot2018local,liu2021elastoplastic,castellanos2021insights,castellanos4015207history}. We extract a structural predictor of local yield strain, called softness, $S$~\cite{cubuk2015identifying,Schoenholz2016}, using neural networks or support-vector machines, demonstrating that our framework can support any local structural measure that predicts rearrangements or local yield stress (e.g.~\cite{richard2020predicting,bapst2020unveiling,paret2020assessing, boattini2021averaging,fan2021predicting,font2022predicting,jung2022predicting}). Following Zhang \textit{et al.}~\cite{zhang2021interplay,zhang2022structuro}, we unravel the interplay between strain, rearrangements and softness, to incorporate softness into structuro-elasto-plasticity (StEP) models. In contrast to EP models, here the local yield strain distribution \emph{emerges} as a collective property.



We develop StEP models for three vastly different systems that each can each be tuned in different ways to exhibit different degrees of strain localization. We demonstrate that the models can be used to gain microscopic insight, opening the door to a quantitative, particle-level approach to engineer advanced structure-property relations in disordered solids. 

\section*{Systems studied}
We examine three systems that differ in dimensionality, temperature, loading condition, and/or interaction potential \cite{sm}. System~I is made of 2D simulated polydisperse circular disks with $1/r_d^{12}$ pairwise repulsions, where $r_d$ is the separation, equilibrated using Monte Carlo swap methods~\cite{ozawa2018random}. Realizations are initially equilibrated at a high temperature $T_a=0.2$ (System~IA) or at a very low temperature of $T_a=0.025$ (System~IB); IA is ductile while IB exhibits strong strain localization and is brittle~\cite{ozawa2018random}. System~II is an experimental granular raft of polydisperse spheres floating at an air-oil interface~\cite{xiao2020strain}. The gravitational capillary length controls the attractive interaction range; in System~IIA the particle diameter $\sigma=1.0 \pm 0.1$~mm is smaller than the capillary length while in IIB the particle diameter $\sigma=3.3 \pm 0.3$~mm exceeds the capillary length. IIB exhibits more strain localization than IIA under tensile strain. Finally, System~III is a three-dimensional polymer nanopillar with chains of coarse-grained particles. The system is simulated at two different temperatures below the glass transition temperature $T_g=0.38$~\cite{lin2019distinguishing, Ivancic2019}, $T=0.3$ (System~IIIA) and $T=0.05$ (IIIB); IIIB exhibits more strain localization under tensile strain than IIIA. For details of all three systems see~\cite{sm}. 

\section*{Softness and yield strain}
For Systems II and III, softness is a weighted sum of local pair and three-body correlation functions obtained using a support vector machine~\cite{Schoenholz2016,zhang2021interplay}. Particles with higher softness tend to have lower yield strains and are therefore more susceptible to rearrangement (Fig.~S1). For details of the definition of softness see~\cite{sm}. The local yield strain $\epsilon_Y$ in poorly annealed jammed packings depends on softness~\cite{zhang2022structuro}, following a Weibull distribution:
\begin{eqnarray}
P(\epsilon_Y,S)=\frac{k}{\lambda}\left(\frac{\epsilon_Y}{\lambda}\right)^{k-1}\exp\left[-(\epsilon_Y/\lambda)^{k}\right]
\label{eq:weibull}
\end{eqnarray}
where $k(S)$ and $\lambda(S)$ characterize the distribution at each softness $S$ (Figs. S4, S5). For System~III, we allow rearrangements to be triggered also by thermal fluctuations~\cite{sm}.  

For System~IB, the standard definition of softness is not nearly as predictive of rearrangements as local yield stress~\cite{richard2020predicting}. We therefore use ResNet to obtain a structural prediction of local yield stress, which we call $Y_p$, see ~\cite{sm} for details (Figs. S2, S3). Softness for this model is simply
\begin{equation}
S=Y_0 -Y_p, \label{eq:SYdef}
\end{equation}
where $Y_0=12.09$ is the mean value of $Y_p$ for the system initially equilibrated at the higher temperature (System~IA), $T_a=0.2$, before quenching to $T=0$. Because local yield strain is a predictor of rearrangements, $S$ serves as a structural predictor of rearrangements~\cite{sm}. The initial value of the mean, $\langle S\rangle$, and the standard deviation, $\sigma_S$, of the softness distribution for each system is shown in Fig.~\ref{fig:flowChart}.

\section*{Structuro-elastoplastic (StEP) models}
\begin{figure*}[tb!]
\begin{minipage}{0.34\textwidth}
\includegraphics[width=\textwidth]{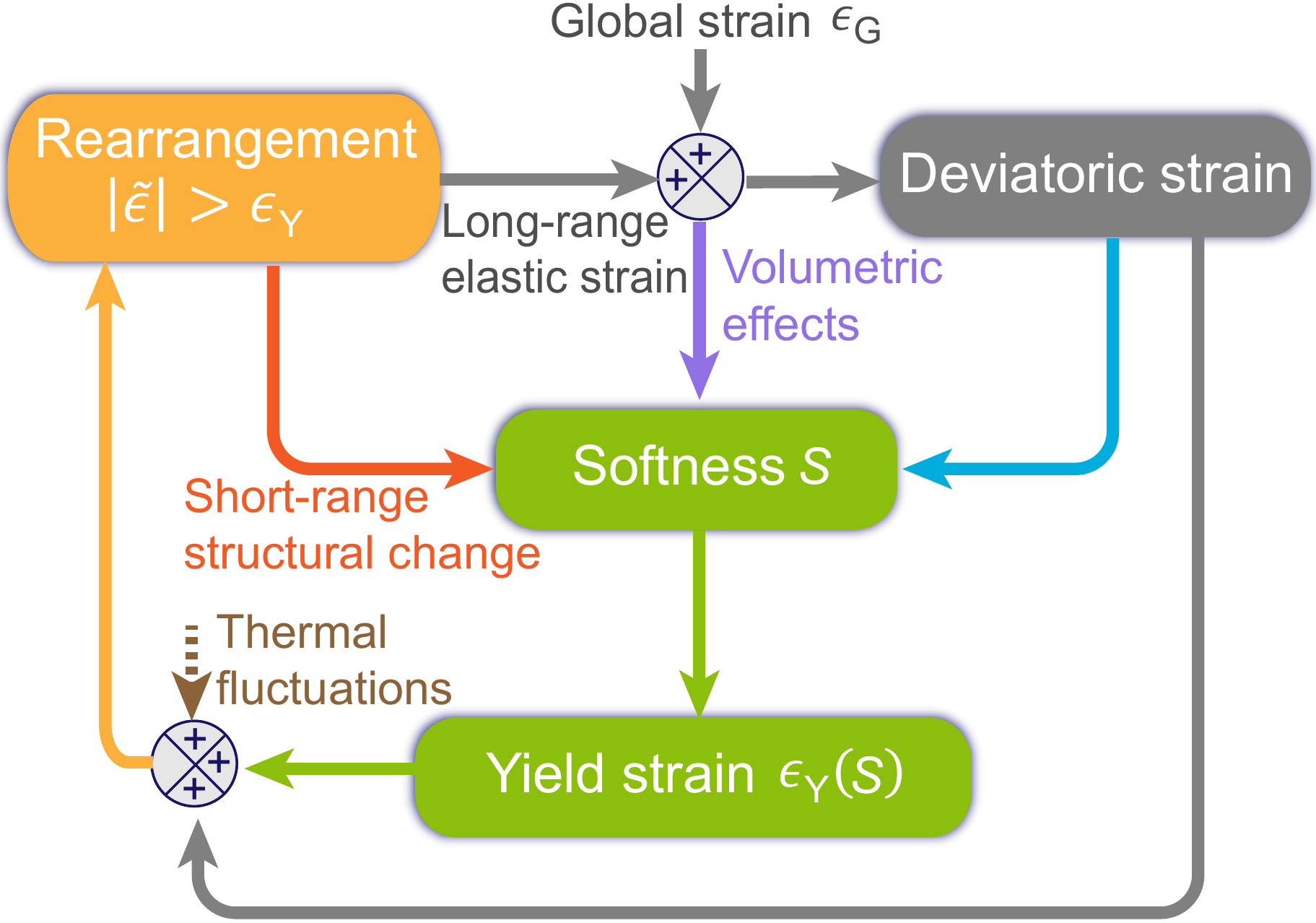}
\end{minipage}
\begin{minipage}{0.66\textwidth}
\resizebox{\textwidth}{!}{
\renewcommand{\arraystretch}{1.45}
\begin{tabular}{ |c|| c| c| c |c |c| c|}
\hline
\hline
\rowcolor{variables}
  Parameters & System~IA & System~IB  & System~IIA & System~IIB & System~IIIA & System~IIIB\\
  \hline
 \rowcolor{softness}
 $\langle S \rangle$ & \color{highlight}{0.00} & \color{highlight}{-2.73}  & -0.09 & -0.04 & 0.575 & 0.085\\ \hline
  \rowcolor{softness}
 $\sigma_S$ & \color{highlight}{3.18} & \color{highlight}{2.28}  & 0.78 & 0.70 & 0.70 & 0.72\\ \hline
  \rowcolor{softness}
 $\langle\epsilon_Y\rangle(S)$ & $(12.09-S)/89$ &  $(12.09-S)/89$ & \color{highlight}{$0.013-0.0006S$} & \color{highlight}{$0.012 -0.00245S$} & \makecell{$\max (0.0484$ \\ $ - 0.102S, 0.0382)$} & $0.0484 - 0.0045S$\\ \hline
  \rowcolor{softness}
 $\epsilon_Y$ distribution &\colorbox{softness}{ \makecell{ Gaussian with \\ $STD=3.08$}} &\colorbox{softness}{  \makecell{ Gaussian with \\ $STD=2.20$}}  &\colorbox{softness}{  \makecell{ Weibull with \\ $k=1.9$}} &\colorbox{softness}{  \makecell{ Weibull with \\ $k=2.1$}} &\colorbox{softness}{  \makecell{ Weibull with \\ $k=2.02$}} &\colorbox{softness}{\makecell{ Weibull with \\ $k=1.84$}}\\ \hline
 \rowcolor{itac}
 $\eta(r)$ & $0.78r^{-2.5}$ & $0.78r^{-2.5}$  & \color{highlight}{$0.1e^{-r/1.4}$} &  \color{highlight}{$0.09e^{-r/0.65}$} & \colorbox{itac}{ \makecell{$0.206e^{-r/1.02} $ \\ $+ 0.059$} }& \colorbox{itac}{ \makecell{ $0.162e^{-r/1.45} $ \\ $ + 0.011$}}\\ \hline
 \rowcolor{itac}
 $c$ & -0.325 & -0.325  & \color{highlight}{0.08} & \color{highlight}{1.20} & 0 & 0\\ \hline
  \rowcolor{zeta}
 $\zeta$ & 244 & 226  & 2.37 & 1.74 & 9.17 & 16.60\\ \hline
 \rowcolor{kappa}
 $\kappa$ & 411 & 411  & 139 & 339 & 13.50 & 11.05\\ \hline
 \rowcolor{cd2min}
 $C_{d2min}$ & $1.0e^{-r/1.88}$ & $1.0e^{-r/1.63}$  & $1.0e^{-r/1.13}$ & $1.0e^{-r/1.12}$ & \color{highlight}{$1.40e^{-r/0.75}$} & \color{highlight}{$0.87e^{-r/1.11}$}\\ \hline
 \hline
\end{tabular}
}
\end{minipage}
\caption{
Schematic and parameters of the StEP model. A strain release (plastic rearrangement event) at a given block changes the softness of nearby blocks, and elastically propagates a long-ranged deviatoric strain field. Softness determines the yield strain for each block. A new rearrangement is triggered if the deviatoric strain exceeds the yield strain. In thermal systems, rearrangements can also be triggered by thermal fluctuations, see supplementary materials for details (Fig.~S11). Structural/elastic/plastic/thermal components of the model are in green/gray/orange/brown, respectively.  Each arrow represents an independently-determined equation, and corresponding parameters shown in the table with matching colors. Key parameters for triggering the ductile-to-brittle transition are highlighted in white and are different for each system.
}
\label{fig:flowChart}
\end{figure*}

We extend the recently-developed 2D and athermal StEP framework~\cite{zhang2022structuro} to the lattice model depicted in Fig.~\ref{fig:flowChart}. We neglect variations of local elastic constants so stress is proportional to elastic strain at the block and system level. For System~I, the softness of site $i$ directly determines the neural-network-predicted local yield stress with $Y_p=Y_0-S_i$ from Eq.~\ref{eq:SYdef} with a distribution determined by that of the actual local yield strain $\epsilon_{Y,i}$~\cite{sm}. For Systems II and III, we assign a yield strain ranking to each site to determine the yield strain from the $S$-dependent distribution of Eq.~\ref{eq:weibull}. This ranking is reassigned after each plastic event at the site~\cite{zhang2022structuro}. Each block's softness is randomly initialized according to the distribution measured in Systems I-IIIAB.

In addition to softness, each block stores a local deviatoric strain tensor, $\tilde \epsilon$.  Each system is driven by a global strain, $\epsilon_G$, uniformly added to all blocks at each time step in the form of simple shear (System~I) or tensile strain (Systems II and III). This uniform strain deforms local structural environments and therefore changes softness. For all of our systems, the change of softness with each strain step is given by $\Delta S_{\text{load}}=\kappa \Delta |\tilde \epsilon|^2$~\cite{zhang2022structuro}, where $\Delta |\tilde \epsilon|^2$ is the increment of $\tilde \epsilon_{\alpha \beta}\tilde \epsilon_{\beta \alpha}$ in each strain step. 

Block $i$ rearranges (undergoes a plastic event) if its strain $|\tilde{\epsilon}|>\epsilon_Y$, the local yield strain. Because StEP models include not only coupling between plasticity and elasticity, but also with structure, a rearrangement at site $i$ affects other sites $j$ via not only an elastic but also a softness kernel. The elastic kernel is described in the supplementary material~\cite{sm}; it is similar to the one commonly used for the $xy$-strain~\cite{budrikis2013avalanche}, but includes more strain components. For block $i$ itself, the elastic strain is converted to plastic strain $\tilde{\epsilon}_p$.

%


%
The softness kernel consists of two main pieces. The first contribution, $\Delta S_{\text{n}}$, is a near-field effect from the change of local structure near a rearrangement, which alters softness of nearby particles directly. For all three systems, we find that this contribution tends to restore $S$ to a value close to the local angular average softness, consistent with Ref.~\cite{zhang2021interplay}. The far-field term, $\Delta S_{\text{f}}$, arises from treating a rearrangement as an Eshelby inclusion that exerts a \textit{far-field} strain that decays as a power law in $r$~\cite{maloney2006amorphous, picard2004elastic,albaret2016mapping}.  This ``elastic facilitation" distorts local structural environments via the volumetric strain, $\tilde\epsilon_\mathrm{vol}$ ~\cite{zhang2021interplay,zhang2022structuro}. Fluctuations around this average behavior are approximated with a Gaussian noise term, $\delta(r)$. Altogether, the softness kernel at distance $r$ from a rearrangement is
\begin{equation}
    \Delta S(r, S)= \underbrace{\eta(r) (\langle S \rangle +c - S) + c^\prime }_{\Delta S_{\text{n}}}+ \underbrace{\zeta\tilde{\epsilon}_{\text{vol}} }_{\Delta S_{\text{f}}}+ \delta(r),
    \label{eq:ds_overall}  
\end{equation}
where the parameters $\eta$, $c$ and $\zeta$ are measured from the corresponding simulations or experiments for each of our systems~\cite{sm}. Here $c^\prime$ is a lattice artifact defined as the average of $-c\langle \eta(r) \rangle$ over all sites; it vanishes in the continuum limit. These parameters for all three systems are extracted from particle simulations or experiments (Figs.~S6-S8~\cite{sm}).

In standard EP models, the lattice size is the rearrangement size. However, the softness kernel in Eq.~\ref{eq:ds_overall} uses $r$ in units of particle size. To avoid rescaling the kernel, we allow
rearrangements in the StEP model to span several blocks~\cite{zhang2022structuro}. The rearrangement size is characterized by the decay length, $\xi$, of the correlation function of particle non-affine displacement~\cite{falk1998dynamics}, $C_{d2min}$, measured for Systems I-III~\cite{sm}, so we allow  blocks at distance $r$ from the rearranging block to release their elastic strain with a probability, $C(r) \propto \exp(-r/\xi)$ (See Figs.~S9-S10~\cite{sm}).

In summary, there are no manually-adjustable parameters in the StEP model. The parameters that appear in the softness kernel, the mean and variance of the softness distribution and the rearrangement size, are all extracted directly from the particle simulations or experiment and are listed in Fig.~\ref{fig:flowChart} for Systems I-III A-B.

\section*{StEP model predictions}

\begin{figure*}
	\centering
	\includegraphics[width=0.95\textwidth]{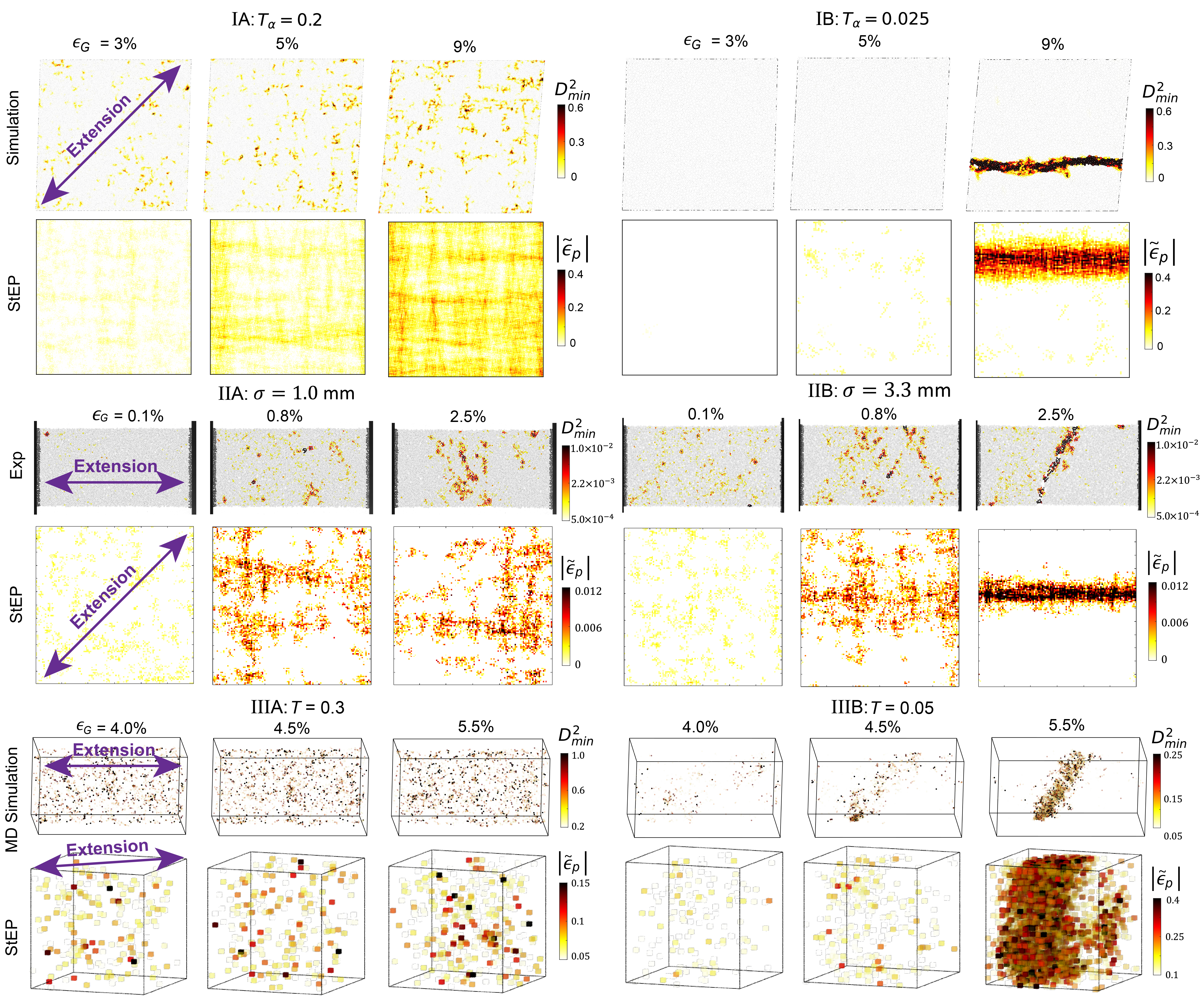}
	\caption{Comparison of spatial plasticity field for particle results and StEP models.  Each column shows the plasticity fields at a different global shear strain $\epsilon_G$. For each system, the first row shows the particle simulation/experiment result, and the second row shows the StEP results. System~I consists of $N$ polydisperse disks at density $\rho=1$ with $1/r_d^{12}$ pairwise repulsion, where $r_d$ is the separation~\cite{ozawa2018random}. System~IA has $N=10000$ disks initially equilibrated at $T_a=0.2$, while System~IB has $N=64000$ disks initially equilibrated at $T_{a}=0.025$, before rapid quenching to $T=0$. System~II consists of $120\sigma\times60\sigma$ granular rafts of polydisperse Styrofoam spheres of mean diameter $\sigma=1.0 \pm 0.1$\,mm (System~IIA) and $\sigma=3.3 \pm 0.3$\,mm (System~IIB), respectively. The spheres float on an air-oil interface and the rafts are subjected to quasi-static extension as indicated. The corresponding StEP models are computed on a 120$\times$120 grid. System~III consists of polymer nanopillars made of $\approx 4 \times 10^4$ bead-spring polymers, with $5$ monomers per chain at temperature $T=0.30$ for System~IIIA and $T=0.05$ for System~IIIB. Each pillar is a cylinder with a length of $100$ bead diameters with periodic boundary conditions, and radius of $25$ bead diameters. The corresponding StEP models are computed on a 21$\times$21$\times$21 grid in units of bead diameter. }
	\label{fig:systems}
\end{figure*}

\begin{figure*}
	\centering
	\includegraphics[width=160mm]{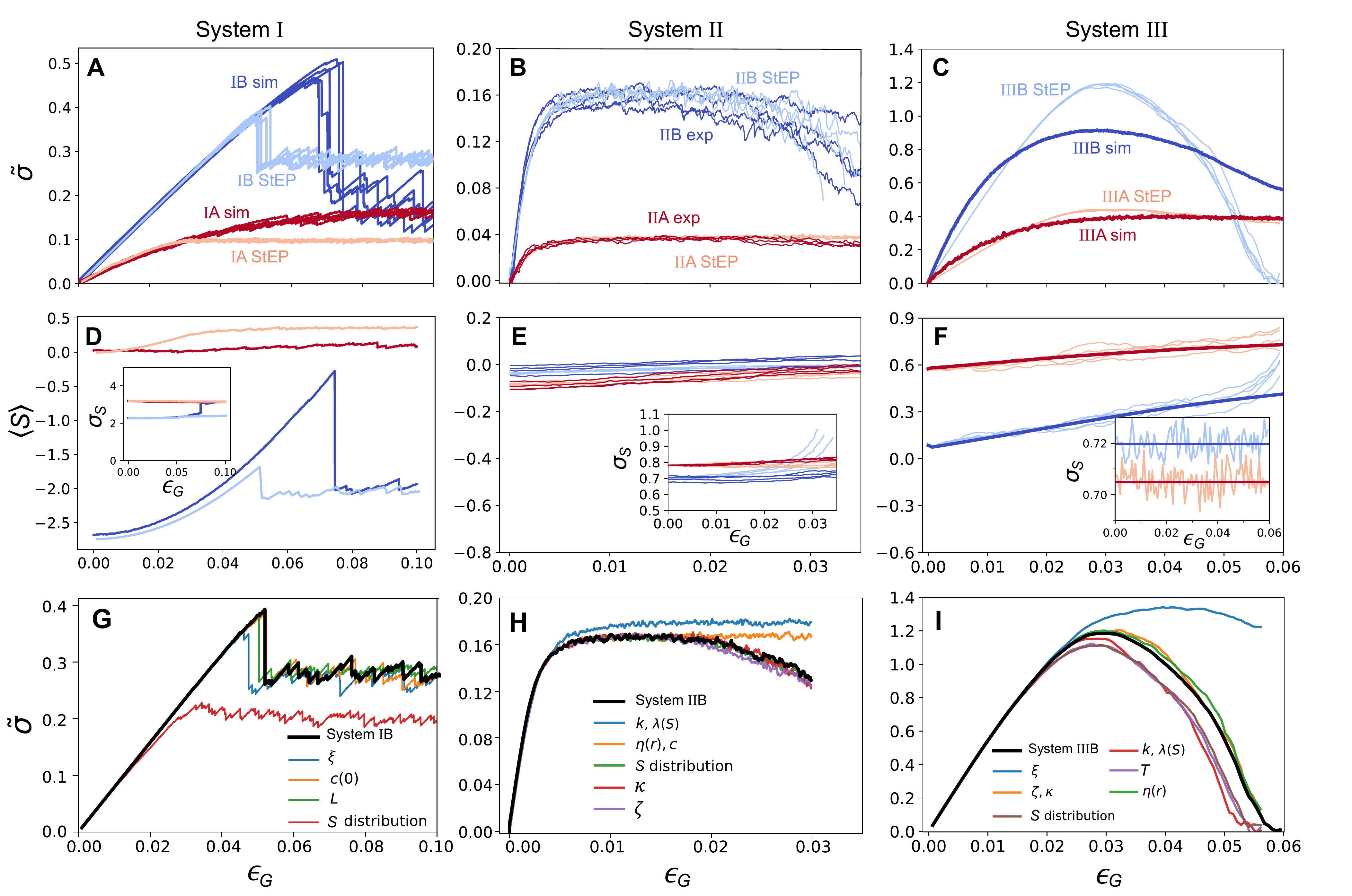}
	\caption{Quantitative comparison between particle simulation/experiment results and StEP models. Cases (I-III)A are denoted by red curves and cases (I-III)B by blue curves. Darker shades correspond to the simulations/experiments while lighter shades correspond to the StEP model. First row:  stress-strain curves; stress is calculated as the product of elastic strain modulus summed over all sites and global modulus measured in the corresponding system. Second row: average softness vs.~strain; insets show standard deviation of the softness distribution vs.~strain. In G-I, the stress-strain curve for the StEP models for Case B is shown as a thick solid black line. For each microscopic factor characterizing the StEP models, the corresponding parameters in Fig.~\ref{fig:flowChart} were varied from their values for Case B to those for Case A. 
	}
	\label{fig:stress-softness}
\end{figure*}

The softness kernel, Eq.~\ref{eq:ds_overall}, successfully captures the difference in ductility in all three model systems. We demonstrate this qualitatively in Fig.~\ref{fig:systems} and quantitatively in Fig.~\ref{fig:stress-softness}, which shows the stress-strain curve as well as the mean and the standard deviation of the softness field for all systems. In the particle systems, plasticity is quantified by the non-affine displacement $D^2_{min}$~\cite{falk1998dynamics}, in units of $\sigma^2$, during a short small applied strain interval $\Delta\epsilon_G$, while in the StEP models, it is the plastic strain $|\tilde{\epsilon}_p|$ during the same strain interval. 
The spatial distribution of the accumulated plasticity is shown at different applied global strains $\epsilon_G$ in Fig.~\ref{fig:systems}. For each system, the difference in strain localization in cases A and B is captured well by the StEP models.

For poorly soft repulsive disks quenched from $T_a=0.2$ to $T=0$, (System~IA),  there is no strain localization and the stress-strain curve shows a smooth yielding process with no stress drops in both StEP-model and particle simulations (Fig.~\ref{fig:stress-softness}A). Spatially correlated rearrangements appear at larger $\epsilon_G$, but there is no system-spanning shear band. For the well-annealed~\cite{ozawa2018random,popovic2018elastoplastic,barlow2020ductile} case quenched from $T_a=0.025$ to $T=0$ (IB), the StEP model captures the sharp shear band that emerges at higher strain, along with accompanying the sharp stress drop (Fig.~\ref{fig:stress-softness}A). Features in the stress-strain curves and the softness statistics (Fig.~\ref{fig:stress-softness}A and D) are captured reasonably well by the StEP models, although the StEP models yield at a lower strain than the particle simulations in the brittle case and the stress drop is less pronounced.

For the experimental granular raft pillars (System~II) in the elastic regime with $\epsilon_G=0.1\%$, small plastic events are distributed throughout the system; this can be seen in both the StEP model and the experiment for Systems~IIA and IIB. As $\epsilon_G$ increases to 0.8\%, transient shear bands at 45$^\circ$ to the principal extension direction are apparent in both the StEP and experimental results. At $\epsilon_G=2.5\%$ , system-spanning shear bands appear in both pillars but with a different morphology. For System~IIA, where the particle interaction range exceeds the particle size, the shear bands are composed of rather sparse rearrangements and are transient in the StEP model; in the experiment the shear bands are locked in the same location due to necking of the pillar. For System~IIB, where the interaction range is smaller than the particle size, the shear band is sharper with much more concentrated plastic events than for IIA, indicating greater strain localization and the location of the shear band is fixed in both the StEP model and experiments. The stress-strain curves, and the evolution of the mean and standard deviation of $S$ with strain are captured remarkably well by the StEP models (Fig.~\ref{fig:stress-softness}B and E). The one exception is the standard deviation of softness after the long-lasting shear band forms in the 3.3~mm particle pillar (System~IIB). This is expected since $S$ is not trained for large fractures in the particle packing.

For the thermal and 3D polymer nanopillar systems (Fig.~\ref{fig:systems}), 
the StEP model still captures the transition from ductile to brittle behaviors. At large $\epsilon_G$, isolated rearrangements are seen for the higher temperature system at $T=0.3$ (IIIA), while strain localization and shear banding occur for the lower temperature case at $T=0.05$ (IIIB) for both the StEP model and simulations. Quantitative comparisons (Fig.~\ref{fig:stress-softness}C and F) show that stress reaches a higher value in the StEP model, but differences between IIIA and IIIB are captured well.


Systems~I-III differ in the size of the constituent particles (and hence importance of temperature), inter-particle interactions, system dimension, importance of friction, and preparation history. Yet in all 6 cases, the StEP models describe the simulations or experiments quantitatively, indicating that they capture the salient microscopic differences in the interplay between elasticity, plasticity and the disordered structure.

\section*{Elucidating how microscopic mechanisms control ductility}

Although preparation history, interaction range and temperature are known to affect strain localization, it is not understood \emph{why} these factors are important. Unlike EP models, StEP models allow us to gain insight by transmuting these factors into the softness kernel and the softness distribution. In particular, the softness kernel is characterized by near-field and far-field contributions in Eq.~\ref{eq:ds_overall}.  By exploring how these two terms, the elastic kernel, the softness distribution and rearrangement size vary between the A and B variations within each system, as well as among our three model systems, we can gain insight into the underlying microscopic mechanisms that control ductility. 

For System~I, we see from the table in Fig.~\ref{fig:flowChart} that Eq.~\ref{eq:ds_overall} is identical for IA and IB. The only striking difference is in the initial mean softness, $\langle S \rangle$. The initial standard deviation of the softness distribution, the relation between $S$ and the yield strain $\epsilon_Y$, and the size of rearrangements are also slightly different. To determine the significance of each of these factors, we start with System~IB (bold black curve in Fig.~\ref{fig:stress-softness}G) and systematically vary each of these parameters (colored curves) one at a time from its value in IB (brittle) to its value in IA (ductile) and assess the effect on the stress-strain curve to see which parameter has the strongest influence. Fig.~\ref{fig:stress-softness}G shows that the only factor that qualitatively affects the stress-strain curve is the initial distribution of $S$. Switching the initial $S$ distribution to its value in the ductile case of System~IA removes the large stress drop in the stress-strain curve, rendering the system ductile. This implies that the only significant difference between Systems IA and IB is the initial softness. This makes sense; the only difference between Systems IA and IB is in the preparation history, which affects the initial value of $S$. Brittle systems obtained by annealing at lower temperature are more stable~\cite{ozawa2018random,berthier2019zero,singh2020brittle}, leading to lower $S$ and hence higher values of the local yield strain from Eq.~\ref{eq:SYdef}. System~I shows that extracting StEP models from particle-level data can lead to correct identification of the microscopic factors controlling strain localization.

For the granular pillars, System~II, the values of nearly all of the parameters differ from System~IIA to IIB. However, by changing them individually (Fig.~\ref{fig:stress-softness}H) from their values for System~IIB (bold black curve) to those for System~IIA, we can see that only two sets of parameters make a significant difference to the stress-strain curve. First, one can flatten the curve by reducing the dependence of the yield strain on softness, characterized by the Weibull exponent $k$ and mode $\lambda(S)$. Second, the near-field softness change caused by rearrangement, \textit{i.e.}, the first term in Eq.~(\ref{eq:ds_overall}), is important. This near-field effect is characterized by the range $\eta (r)$ and offset $c$. System~IIA has a larger decay length of $\eta (r)$ and a smaller value of $c$. In other words, rearrangements in System~IIB, which has a shorter particle interaction range, alter local yield strain more strongly over a shorter distance.  The range of the near-field softness change is comparable to the interaction range. 

\newcommand{\dtwomincorr}[0]{\ensuremath{\xi}}
The values of most of the parameters for the polymer nanopillar StEP models differ from System~IIIA to IIIB (table in Fig.~\ref{fig:flowChart}), but we change them one by one from their values for System~IIIB to their values for System~IIIA in Fig.~\ref{fig:stress-softness}I. None of them affect the stress-strain curve appreciably except rearrangement size, $\dtwomincorr$. The rearrangements in the low temperature system, System~IIIB, have a larger size $\dtwomincorr$, allowing for more facilitation.

\section*{Discussion}

For three systems, we have shown that structuro-elasto-plasticity models provide physical insight into the microscopic mechanisms that govern their ductility.  The systems were designed to be very different, and we find that the microscopic mechanisms underlying the degree of strain localization are also different. For simulated glasses with different preparation histories, we find that it is the initial distribution of softness, and hence of local yield stress, that controls ductility. For experimental granular pillars with different interaction ranges, it is the sensitivity of local yield strain on local structure and the near-field change of softness due to rearrangements that change, with a larger change of local yield stress occurring over a shorter range in systems with higher strain localization. 
Finally, for the simulated polymer glass nanopillars, we find that the rearrangement size is larger in cooler systems, allowing for more facilitation and leading to greater strain localization. These insights would not be possible if we had not been able to connect local structure with rearrangement propensity. The introduction of the softness field is critical.

Our results raise the possibility that StEP models could be used to design materials with desired ductility by optimizing over multiple controllable factors. This would be a substantial improvement over empirical approaches. This hope does not seem unrealistic; softness has been proved to be highly predictive of rearrangements in a wide range of disordered solids~\cite{cubuk2017structure, yang2022understanding}. Moreover, the StEP model framework is highly adaptable, accommodating any structural predictor of rearrangements, as we have shown by introducing a structural predictor of local yield stress obtained using image classification methods instead of the more standard predictor of rearrangements based on a weighted average of two- and three-point structure functions.

It is straightforward to extend StEP models to include dynamics by using time-dependent softness and elastic kernels~\cite{liu2021elastoplastic}, in order to capture rheology. Like EP models, StEP models can also readily be extended to include local elastic moduli and other effects~\cite{nicolas2018deformation}.


\section*{Acknowledgements}

We thank Misaki Ozawa for providing the initial soft-disk configurations used in this study. \textbf{Funding}: This work was supported by the National Science Foundation through grant MRSEC/DMR-1720530 (D.J.D., R.A.R., H.X., E.Y., G.Z.), and the Simons Foundation via the ``Cracking the glass problem" collaboration \#45945 (S.A.R., A.J.L.) and Investigator Award \#327939 (A.J.L.). The Extreme Science and Engineering Discovery Environment (XSEDE), supported by National Science Foundation grant number ACI-1548562. This work used the XSEDE Stampede2 at the Texas Advanced Computing Center through allocation TG-DMR150034 (E.Y., R.J.S.I, R.A.R.). \textbf{Author contributions}: G.Z. performed simulations for System~I, H.X. performed experiments for System~II, R.J.S.I. performed simulations for System~III, G.Z., H.X., E.Y., R.J.S.I., S.A.R. performed analysis to data and developed the StEP model predictions. G.Z., H.X., R.J.S.I, E.Y. drafted the original manuscript, all participated in reviewing and editing, A.J.L, D.J.D, R.A.R. were responsible for conceptualization, funding acquisition, and supervision. \textbf{Competing interests}: the authors have no competing interests. \textbf{Materials availability}: Materials are available upon request to ajliu@physics.upenn.edu.

\nocite{ozawa2020role}
\nocite{zagoruyko2016wide}
\nocite{nicolas2013mesoscopic}
\nocite{Yang_2022}

\bibliography{scibib}
\bibliographystyle{Science}

\end{document}



\baselineskip24pt


\maketitle


\section{Particle systems}
\subsection{System I: Soft repulsive disks}

System I is a computer-simulated 2D polydisperse soft-disk system detailed in Ref.~(47). We received configurations equilibrated at several different initial temperatures using the swap Monte Carlo algorithm from the authors of Ref.~(47). Note that the configurations prepared initially at $T=0.025$, used for System IB, may not have been completely equilibrated before the quench to $T=0$, but this does not affect interpretation of the results.

After potential-energy minimization to zero temperature, $T=0$, these configurations are sheared quasistatically by repeatedly applying a small simple shear strain of $\delta \epsilon = 10^{-5}$ and minimizing the total potential energy, until the total strain reaches $\epsilon_{\text{end}}=0.1$. Energy minimization is carried out using the steepest descent algorithm with a very conservative step size to simulate overdamped dynamics, as described in the supplementary material of Ref.~(36). During the minimization, we study intermediate configurations using the same protocol as Ref.~(36). These intermediate configurations allow us to investigate the interplay of relevant quantities during an avalanche.

\subsection{System II: Granular rafts}

System II is an experimental granular raft made of a disordered monolayer of polydisperse granular particles floating at an air-oil interface~(9). Mineral oil was used with a density of $\rho_{\text{oil}}=$870 $\pm$ 10\,kg/m$^3$, and the particles are made of Styrofoam with a density of 15\,kg/m$^3$. Systems IIA and IIB correspond to rafts with different particle size distributions, with average particle sizes of 1.0\,$\pm$0.1\,mm for IIA and 3.3\,$\pm$\,0.3\,mm for IIB. Capillary attractions exist between nearby particles, with a characteristic interaction range set by the capillary length of the oil, $l_c=\sqrt{\gamma_{\text{oil}}/\rho_{\text{oil}} g}=1.8$\,mm, where $\gamma_{\text{oil}}=27.4$\,dyn\,cm$^{-1}$, and $g=$9.8\, m/s$^2$. 
For each experiment, particles were assembled into a disordered monolayer (pillar) with a rectangular shape~(9). Quasi-static tensile tests with a strain rate on the order of $10^{-5}$\,s$^{-1}$ were then conducted and all particle positions and the global tensile force were tracked throughout the experiments. For extracting microscopic information and constructing StEP models, results from 50 experiments~(9) with $80\sigma\times40\sigma$ sized pillars for each particle size were used. For comparing with StEP model results, results from 12 experiments of larger pillars of size $120\sigma\times60\sigma$ for each particle size were used for better statistics.

\subsection{System III: Polymer nanopillars}

%
Using LAMMPS, we simulate bead-spring polymer nanopillars with $N = 5$ monomers per chain as detailed in Ref.~(39). We bond the monomers using a stiff harmonic potential $U_{i,j}^b = \frac{k}{2} \left( r_{i,j} - \sigma \right)^2$ where $r_{i,j}$ is the distance between monomers $i$ and $j$ and $k = 2000 \epsilon/\sigma$. We take the nonbonded interactions as a modified Lennard-Jones potential $U_{i,j}^{nb} = 4 \epsilon \left( \left( \frac{\sigma'}{r_{i,j}-\Delta} \right)^{12} - \left( \frac{\sigma'}{r_{i,j}-\Delta} \right)^{6} \right)$ where $\sigma' = (1-3/2^{13/6}) \sigma$ and $\Delta = 3\sigma/4$. This modification increases the curvature of the potential while maintaining the minima at $r_{\text{min}} = 2^{1/6} \sigma$, causing more brittle failure~(8). For all of our simulations, we use a timestep of $0.000663652$ to compensate for this additional curvature. We thermalize our simulations within a cylindrical, harmonic confining wall at $T=0.5$ that we fix to ensure the density of the monomers within the wall is $\rho = 0.3$. We then cool our simulation at a rate of $5 \times 10^{-4}$ past the glass transition temperature of $T_{g} \approx 0.38$ to $T=0.30$ (Case A) and $T=0.05$ (Case B) causing the density to increase to $\rho \approx 1.0$ at the lower temperature. We then deform the nanopillars at a true strain rate of $\dot{\epsilon} = 10^{-4}$. We repeat this procedure for $50$ replicas. We output monomer positions every $10000$ timesteps. 

\section{Softness training}
\subsection{Softness training using an SVM}

\begin{figure}[t]
\centering
\includegraphics[width=0.45\linewidth]{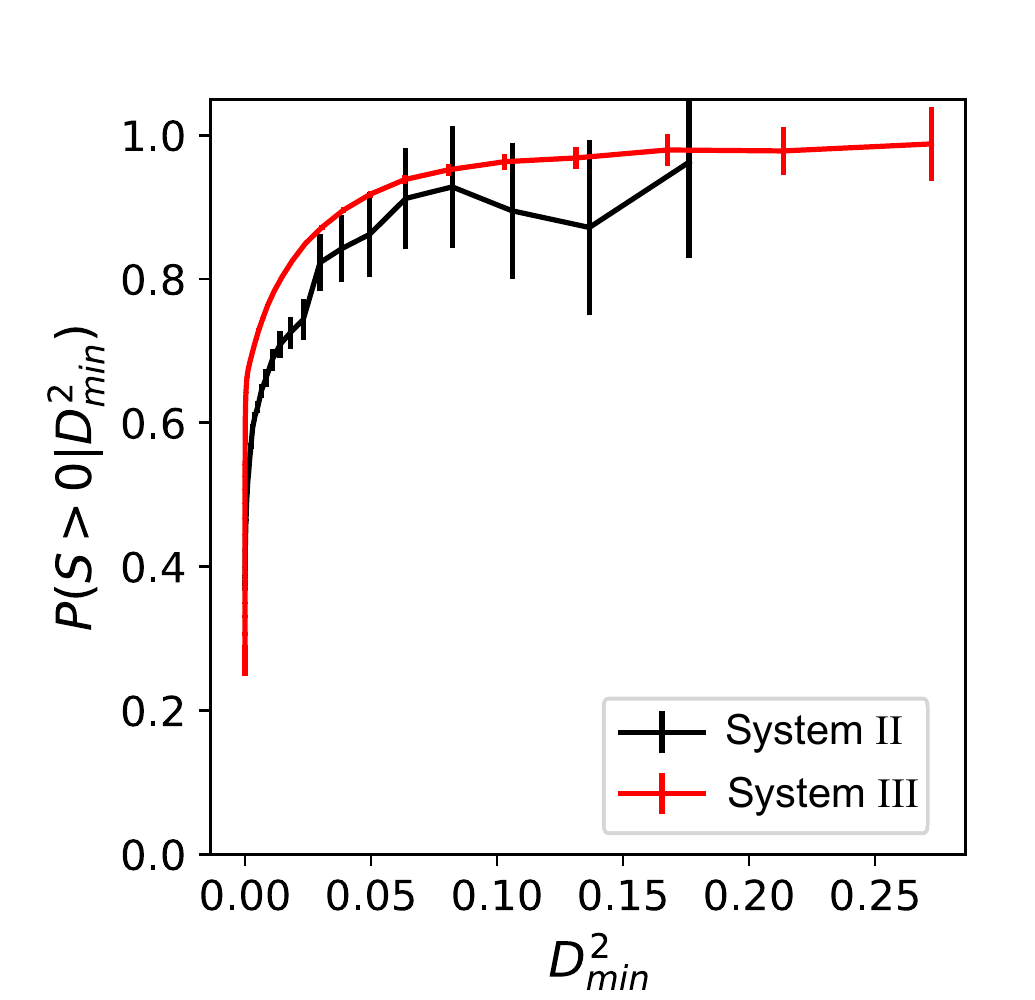}
\caption{Probability that a particle of a given $D^2_{\text{min}}$ is soft for the granular raft experiments (System II) and polymer nanopillar simulations (System III). Here we scale $D^2_{\text{min}}$ by the particle radii $\sigma$ in each case.}
\label{fig:P_soft_given_D2min}
\end{figure}

We train the softness field for Systems II and III similarly to Refs.~(27) and~(28). For System III we analyze the time-averaged monomer positions during these trajectories, averaging every $5$ timesteps in the $500$ timesteps before each frame. We first extract a set of rearranging and non-rearranging particles from the early stage of the deformation (pre-shear band formation). To differentiate between these groups, we calculate a $D^2_{\text{min}}$ field~(19) in which we take the strain between frames to be commensurate with the amount of strain for a particle to complete a rearrangement ($\epsilon_G \approx 2.0\times10^{-3}$ for System II and $\epsilon_G \approx 6.6\times10^{-4}$ for System III) and take the cutoff radius ($R_{c} = 1.75 \sigma$) to include the first shell of neighbors. We consider this field's local maxima and minima to be the rearranging and non-rearranging particles. We next encode the local structure around these training examples as a high-dimensional vector ($\vec{G} \in \mathbb{R}^N$) 
which we describe in detail elsewhere~(28). Because large longitudinal strains cause significant changes in local density at later strains, we also consider these structural features scaled by deviations of the local packing density from the average density $\left(\rho - \langle \rho \rangle \right)/\langle \rho \rangle$. We append this scaled vector to the original vector to fully describe the local structure of our training examples, $\vec{G}^\frown \left(\left(\rho - \langle \rho \rangle \right)/\langle \rho \rangle \vec{G} \right)$. We use these appended structural descriptors in $\mathbb{R}^{2N}$ as the input for a linear support vector machine (SVM) classification to calculate the hyperplane that best separates rearranging particles from non-rearranging particles. We compute softness as the signed distance from the hyperplane to the data point in the appended high-dimensional structural descriptor space. We exclude particles on the exterior of the pillars from training and testing for simplicity. We use the same softness field for both particle sizes in the granular raft experiment (System II) by rescaling by the average particle radius. To validate this model, we consider the probability of a particle being soft at a given $D^2_{\text{min}}$ $P(S>0|D^2_{\text{min}})$ for a set of pillars independent from the pillars that were use to construct the test set, Fig.~\ref{fig:P_soft_given_D2min}. This function has a strong monotonically increasing dependence on $D^2_{\text{min}}$ that plateaus near $1$ suggesting a strong correlation between being soft and obtaining a large dynamic event. 


\subsection{Softness training using a neural network}

The standard definition of softness used for Systems II and III correlates only very weakly with rearrangements for System IB.  For System I we therefore use a neural network instead of a support vector machine to define softness.

\begin{figure}[t]
\centering
\includegraphics[width=0.35\linewidth, trim={0.92cm 0.65cm 0 0},clip]{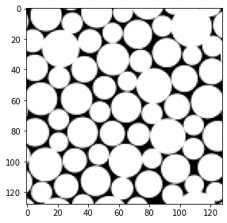}
\caption{An example input of the neural network for System I.}
\label{fig:neuralNetworkExampleInput}
\end{figure}

We train a 17-layer residual convolutional neural network to predict the local yield stress of each particle. The architecture is similar to the ``wide residual network''~(48), except for the following differences. First, the input is changed to 128-by-128 grayscale (single channel) images, cropped from an image of the entire configuration, which has a resolution of 2048-by-2048 for configurations with $N=10000$ particles and 4096-by-4096 for configurations with $N=64000$ particles. One example of such an image is presented in Fig.~\ref{fig:neuralNetworkExampleInput}. Second, we used four groups of convolutions instead of three, since our input image is larger. Each group contains two ResNet blocks, as in~(48). Third, we used the ``basic'' version of ResNet blocks in Fig.~1 of Ref.~(48). While Ref.~(48) found that wider versions are better for image classification tasks, we find that they provide little improvement for our tasks. Fourth, We removed the last global-average pooling layer and softmax layer of the original neural network, because they are only suitable for image classification tasks. These layers are replaced with a fully-connected layer with a single neuron outputting a predicted local yield stress. Our loss function is the squared difference between the predicted and the actual local yield stress. We calculate the actual local yield stress using the procedure detailed in Ref.~(37). Last, we impose an L2 regularizer with regularization parameter $0.2$ on all weights and biases of the neural network. We also augment the training data by randomly flipping it in both the horizontal and the vertical directions.

The neural network was trained on the configurations after equilibration at three different temperatures but before performing quasistatic shear. Both the training dataset and the test dataset are derived from five independent configurations with  $N=10000$ particles equilibrated at $T_a=0.025$, one configuration with $N=64000$ and $T_a=0.1$, and one configuration with $N=64000$ and $T_a=0.2$. For each configuration, we calculate the local yield stress of every particle using the protocol detailed in Ref.~(37), and asked the neural network to predict it. The neural network was trained in 40 epochs with batch size 16. We used Adam minimizer for training with a learning rate that decays exponentially from $3\times10^{-4}$ to $3\times10^{-6}$. After training, the coefficient of determination on the test set is $R^2=0.5809$, i.e., 58.09\% of the variance in the test data is captured by the prediction. Let the predicted local yield stress for a particle be $Y_p$, we then define softness as $S=\langle Y_p \rangle -Y_p$, where $\langle Y_p \rangle=12.09$ is the average of $Y_p$ at $T_a=0.2$.

\section{Constructing StEP models from simulations and experiments}

\subsection{Elastic kernel functions in 2D and 3D}


The elastic kernel function is the strain change in other sites caused by a unit strain release at the origin. Our model in 2D explicitly includes two strain components: $\epsilon_{xy}$ and $\epsilon_{xx-yy}=(\epsilon_{xx}-\epsilon_{yy})/2$, and therefore uses four elastic kernels: a $\epsilon_{xy}$ to $\epsilon_{xy}$ kernel, a $\epsilon_{xx-yy}$ to $\epsilon_{xx-yy}$ kernel, and two cross-term kernels. Each kernel was derived using the ``Fourier discretized'' method~(40). Specifically, we start from analytical Fourier-space kernel functions, discretize on a lattice with the same size as the real-space system, and then perform an inverse discrete Fourier transform to obtain the kernel in the real space. The reason for this process is that digitization in the Fourier space is equivalent to applying periodic boundary conditions in the real space.

The analytical expression for the Fourier-space kernel in 2D is obtained from Ref.~(49). It is
\begin{equation}
\begin{pmatrix}
\epsilon_{xx-yy}(\mathbf q)\\
\epsilon_{xy}(\mathbf q)
\end{pmatrix}
=\frac{1}{|\mathbf q|^4}
\begin{pmatrix}
-(q_x^2-q_y^2)^2 & -2q_xq_y(q_x^2-q_y^2)\\
-2q_xq_y(q_x^2-q_y^2) & -4q_x^2q_y^2
\end{pmatrix}
\begin{pmatrix}
\epsilon_{p, xx-yy}\\
\epsilon_{p, xy}
\end{pmatrix}
\end{equation}
where $\epsilon_{xx-yy}(\mathbf q)$ and $\epsilon_{xy}(\mathbf q)$ are the elastic strain changes, $\mathbf q=(q_x, q_y)$ is the Fourier-space coordinates, and $\epsilon_{p, xx-yy}$ and $\epsilon_{p, xy}$ are the strain releases at the origin.

In 3D, we have five strain components: $\epsilon_{xx}$, $\epsilon_{xy}$, $\epsilon_{xz}$, $\epsilon_{yy}$, and $\epsilon_{yz}$. Thus, there are $25$ elastic kernels in total. Similar as 2D cases, each kernel can be derived using the ``Fourier discretized'' method and the analytical expression is provided below:

\begin{equation}
\begin{pmatrix}
\epsilon_{xx}(\mathbf q)\\
\epsilon_{xy}(\mathbf q)\\
\epsilon_{xz}(\mathbf q)\\
\epsilon_{yy}(\mathbf q)\\
\epsilon_{yz}(\mathbf q)
\end{pmatrix}
=\frac{1}{|\mathbf q|^4}
\begin{pmatrix}
G_{11} & G_{12} & G_{13} & G_{14} & G_{15} \\
G_{21} & G_{22} & G_{23} & G_{24} & G_{25} \\
G_{31} & G_{32} & G_{33} & G_{34} & G_{35} \\
G_{41} & G_{42} & G_{43} & G_{44} & G_{45} \\
G_{51} & G_{52} & G_{53} & G_{54} & G_{55} 
\end{pmatrix}
\begin{pmatrix}
\epsilon_{p, xx}\\
\epsilon_{p, xy}\\
\epsilon_{p, xz}\\
\epsilon_{p, yy}\\
\epsilon_{p, yz}
\end{pmatrix}
\end{equation}

\begin{multicols}{2}
\begin{equation}
\begin{pmatrix}
G_{11}\\
G_{21}\\
G_{31}\\
G_{41}\\
G_{51}\\
\end{pmatrix}
=
\begin{pmatrix}
2q_x^2(q^2-q_x^2+q_z^2) - q^4 \\ 
q_xq_y(q^2 - 2q_x^2+2q_z^2) \\
2q_xq_z(-q_x^2+q_z^2) \\
2q_y^2(-q_x^2 + q_z^2) \\
q_yq_z(-q^2-2q_x^2+2q_z^2)
\end{pmatrix}
\end{equation}

\begin{equation}
\begin{pmatrix}
G_{12}\\
G_{22}\\
G_{32}\\
G_{42}\\
G_{52}\\
\end{pmatrix}
=
\begin{pmatrix}
2q_xq_y(q^2-2q_x^2) \\ 
-4q_x^2q_y^2 - q^2q_z^2 \\
q_yq_z(q^2 - 4q_x^2) \\
2q_xq_y(q^2 - 2q_y^2) \\
q_xq_z(q^2-4q_y^2)
\end{pmatrix}
\end{equation}
\end{multicols}

\begin{multicols}{2}
\begin{equation}
\begin{pmatrix}
G_{13}\\
G_{23}\\
G_{33}\\
G_{43}\\
G_{53}\\
\end{pmatrix}
=
\begin{pmatrix}
2 q_x q_z(q^2 - 2q_x^2) \\ 
q_y q_z(q^2 - 4q_x^2) \\
-4 q_x^2 q_z^2 - q^2 q_y^2 \\
-4 q_x q_y^2 q_z \\
q_x q_y(q^2 - 4 q_z^2)
\end{pmatrix}
\end{equation}

\begin{equation}
\begin{pmatrix}
G_{14}\\
G_{24}\\
G_{34}\\
G_{44}\\
G_{54}\\
\end{pmatrix}
=
\begin{pmatrix}
2 q_x^2 (-q_y^2 + q_z^2) \\ 
q_xq_y(q^2 - 2q_y^2 + 2q_z^2) \\
q_xq_z(-q^2 - 2q_y^2 + 2q_z^2) \\
2 q_y^2 (q^2 - q_y^2 + q_z^2) - q^4 \\
2 q_yq_z (-q_y^2 + q_z^2)
\end{pmatrix}
\end{equation}

\end{multicols}

\begin{equation}
\begin{pmatrix}
G_{15}\\
G_{25}\\
G_{35}\\
G_{45}\\
G_{55}\\
\end{pmatrix}
=
\begin{pmatrix}
-4 q_x^2 q_y q_z \\ 
q_x q_z (q^2 - 4 q_y^2) \\
q_x q_y (q^2 - 4 q_z^2) \\
2 q_y q_z (q^2 - 2 q_y^2) \\
-4 q_y^2 q_z^2 - q^2 q_x^2
\end{pmatrix}
\end{equation}
where $\epsilon_{xx}(\mathbf q)$, $\epsilon_{xy}(\mathbf q)$, $\epsilon_{xz}(\mathbf q)$, $\epsilon_{yy}(\mathbf q)$, and $\epsilon_{yz}(\mathbf q)$ are the elastic strain changes, $\mathbf q=(q_x, q_y, q_z)$ is the Fourier-space coordinates, and $\epsilon_{p, xx}$, $\epsilon_{p, xy}$, $\epsilon_{p, xz}$, $\epsilon_{p, yy}$, and $\epsilon_{p, yz}$ are the strain releases at the origin.

\subsection{Obtaining yield strain distributions for Systems I-III}

\begin{figure}[t]
\centering
\includegraphics[width=1.0\linewidth]{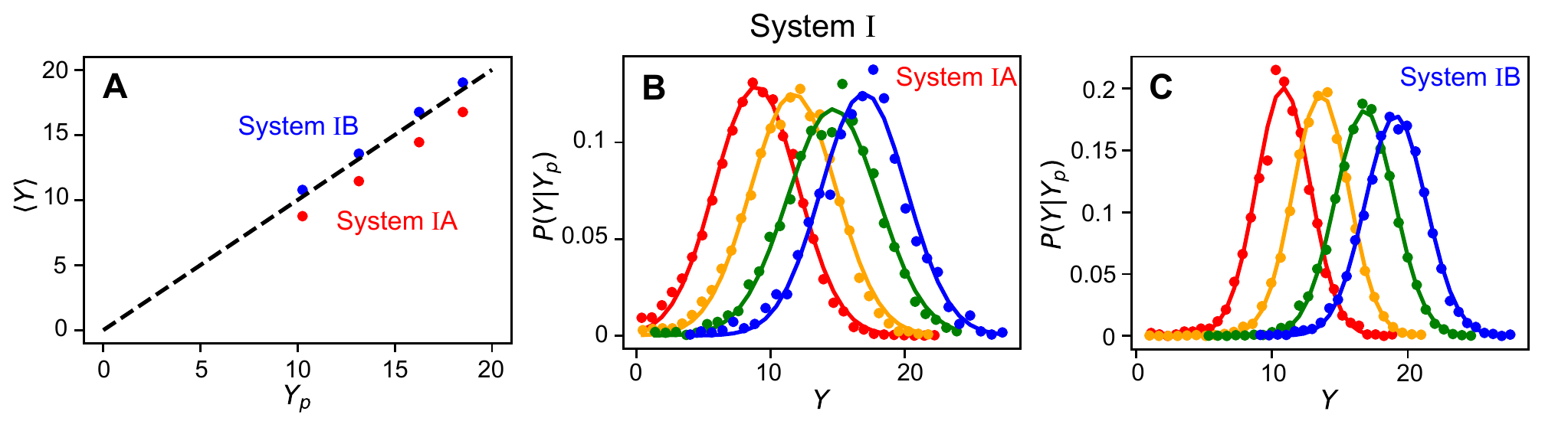}
\caption{Measurement of local yield stress for System I. (A) Mean actual local yield stress $\langle Y \rangle$ versus predicted local yield stress $Y_p$ for System I. Black dotted line represents $Y_p=Y$, i.e., perfect predictions from the neural network. (B) The distribution of the actual local yield stress $Y$ for particles with predicted local stress $Y_p$ in the range (red) $10<Y_p<10.5$, (yellow) $13<Y_p<13.3$, (green) $16<Y_p<16.5$, and (blue) $18<Y_p<19$, respectively, for the 2D soft-disk systems equilibrated at $T_a=0.2$ (System IA). Dots are histograms, and lines are their Gaussian fits. (C) Same as (B), except for $T_a=0.025$ (System IB).}
\label{fig:localyieldstrain2}
\end{figure}

\begin{figure}[t]
\centering
\includegraphics[width=1.0\linewidth]{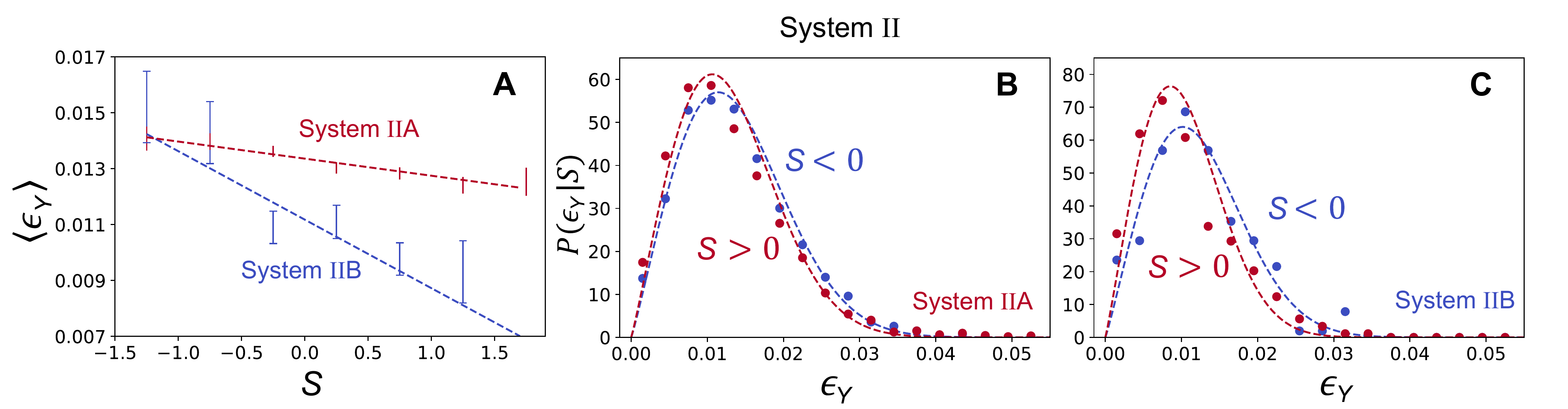}
\caption{Measurement of local yield strain for System II, granular rafts. (A) Mean yield strain vs. softness for 1.0\,mm rafts (System IIA, red) and 3.3\,mm rafts (System IIB, blue). The dotted lines are corresponding linear fittings, $\langle\epsilon_Y\rangle = 0.0134 - 0.000611 S$, for IIA, and, $\langle\epsilon_Y\rangle = 0.0118 - 0.00245 S$, for IIB. (B) and (C) distribution of yield strain for specific softness for the 1\,mm (System IIA) and 3\,mm (System IIB) rafts, respectively. The dotted curves are Weibull distribution with the mean $\langle\epsilon_Y\rangle$ and the shape parameter $k=1.9$ for System IIA rafts and $k=2.1$ for System IIB.}
\label{fig:localyieldstrain}
\end{figure}

The local yield strain is measured in two different ways. For the soft disk simulations of System I, a ``frozen matrix method'' was used~(37). As detailed in Ref.~(37), the method is similar to Ref.~(23) except that we do not project the yield strains in different directions to the global-shear direction. For the experimental granular rafts (System II) and the polymer nanopillars (System III), a 'rewinding' method was used. For each rearrangement, its yield strain $\epsilon_Y$ is measured as the deviatoric elastic strain accumulated between $t_0$ and $t_{re}$, where $t_0$ is the time corresponding to the beginning of the deformation or the end of the previous rearrangement at the same location, and $t_{re}$ is the beginning of the current rearrangement. 

For System I, 2D soft disks, the softness $S=\langle Y_p \rangle -Y_p$ is defined as a function of the neural network's predicted local yield stress $Y_p$. Therefore, it is more straightforward to compare $Y$ and $Y_p$. In general, $Y_p$ and $Y$ are close but not exactly the same, since the neural network is not perfectly accurate. In Fig.~\ref{fig:localyieldstrain2}B and C we plot the distribution of $Y$ for particles with $Y_p$ residing in different bins. We find that the distribution shapes are closer to Gaussian rather than Weibull functions. In Fig.~\ref{fig:localyieldstrain2}A we show that the mean of $Y$ is always very close to $Y_p$. We also found that the standard deviation of the distributions does not vary significantly across bins. Based on these observations, in StEP simulations we let the distribution of $Y$ be a Gaussian with mean equal to $Y_p$, and standard deviation equal to 3.1 for $T_a=0.2$ (System IA) and 2.2 for $T_a=0.025$ (System IB). 
The local yield {\it strain} $\epsilon_Y$ can then be calculated by dividing $Y$ by the average shear modulus, $G=89$.

\begin{figure}[t]
\centering
\includegraphics[width=1.0\linewidth]{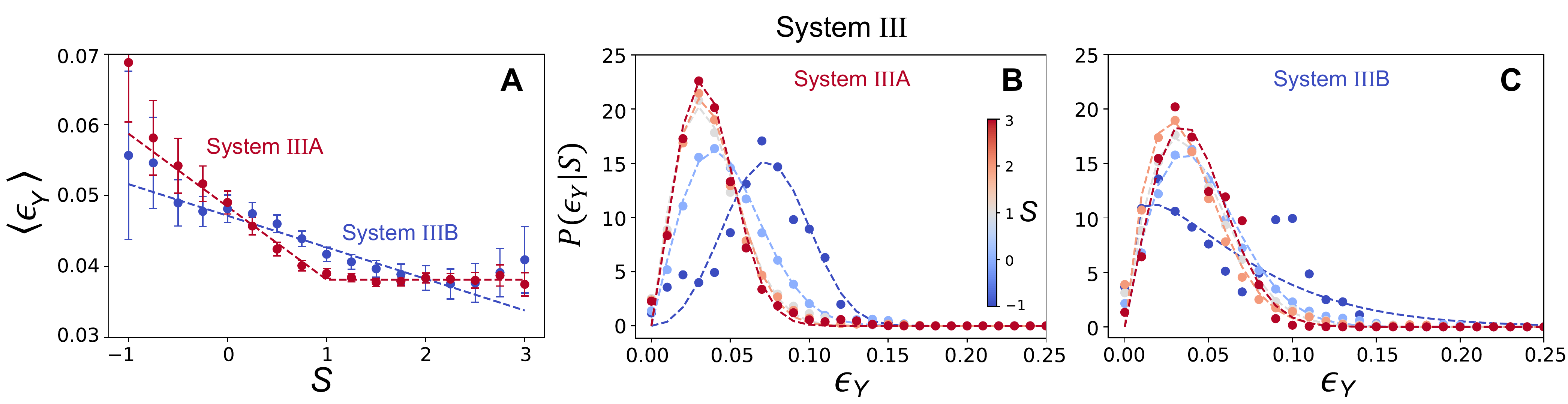}
\caption{Measurement of local yield strain in System III, polymer nanopillars. (A) Mean yield strain vs. softness for and $T=0.30$ pillars (IIIA, red) and $T=0.05$ pillars (IIIB, blue). The dash lines are corresponding linear fittings, $\langle\epsilon_Y\rangle = \max (0.0484 - 0.102S,  0.0382)$, for IIIA, and,  $\langle\epsilon_Y\rangle = 0.0472 - 0.0045S$, for IIIB. (B) and (C) are Distribution of yield strain for specific softnesses for (B) IIIA and (C) IIIB. The dashed curves are Weibull distributions with the mean $\langle\epsilon_Y\rangle = \max (0.0484 - 0.102S,  0.0382)$ and the shape parameter $k=2.02$ for IIIA, and $\langle\epsilon_Y\rangle = 0.0472 - 0.0045S$ and $k=1.84$ for IIIB. The color gradient represents the softness gradient.}
\label{fig:localyieldstrain3}
\end{figure}

Figure~\ref{fig:localyieldstrain} shows the relation between the local yield strain and softness for System II, the experimental granular rafts. The measured local yield strain is binned according to the local softness before rearrangement and the binned-averaged value, $\langle \epsilon_Y\rangle$, is plotted against $S$ in Fig.~\ref{fig:localyieldstrain}A for granular materials. Examples of measured distributions for specific softness ranges are shown in Fig.~\ref{fig:localyieldstrain}B and C. Results for polymer nanopillars, System III, are shown in Fig.~\ref{fig:localyieldstrain3}.

\subsection{Near-field structural contribution to softness kernel}

\begin{figure}[t]
\centering
\includegraphics[width=0.8\linewidth]{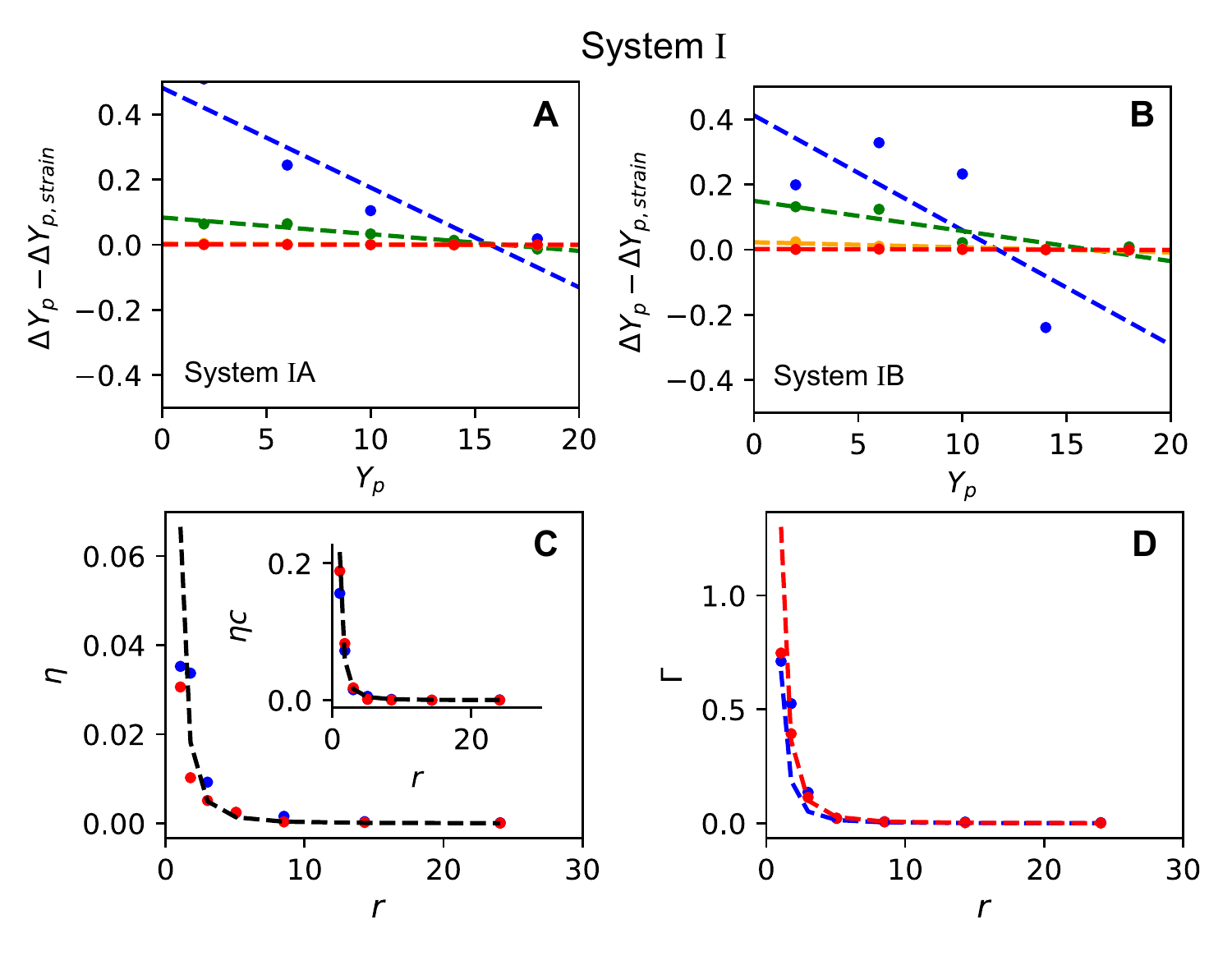}
\caption{Near-field structural contribution to the change of the predicted local yield stress, $\Delta Y_p$, due to rearrangements for 2D soft repulsive disks (System I), after subtracting the strain contribution $\Delta Y_{p, strain}$. (A) restoring effect for $T_a=0.2$ (System IA), for neighbors of rearrangers within distance $r<1.34$ (blue), $2.25<r<3.78$ (green), $6.35<r<10.68$ (orange), and $17.96<r<30.20$ (red); (B) restoring effect for $T_a=0.025$ (System IB), colors have the same meanings as (A); (C) restoring coefficient for systems IA (red) and IB (blue); and (D) variance of softness change, colors have the same meanings as (C). In (C), we found that two common equations, $\eta(r)=0.0779r^{-2.5}$ and $\eta(r)c=0.253r^{-2.5}$, can roughly fit $\eta(r)$ and $\eta c(r)$ for both temperatures, and the fit is therefore indicated by a black dotted line.}
\label{fig:restore_softDisk}
\end{figure}

\begin{figure}[t]
\centering
\includegraphics[width=0.8\linewidth]{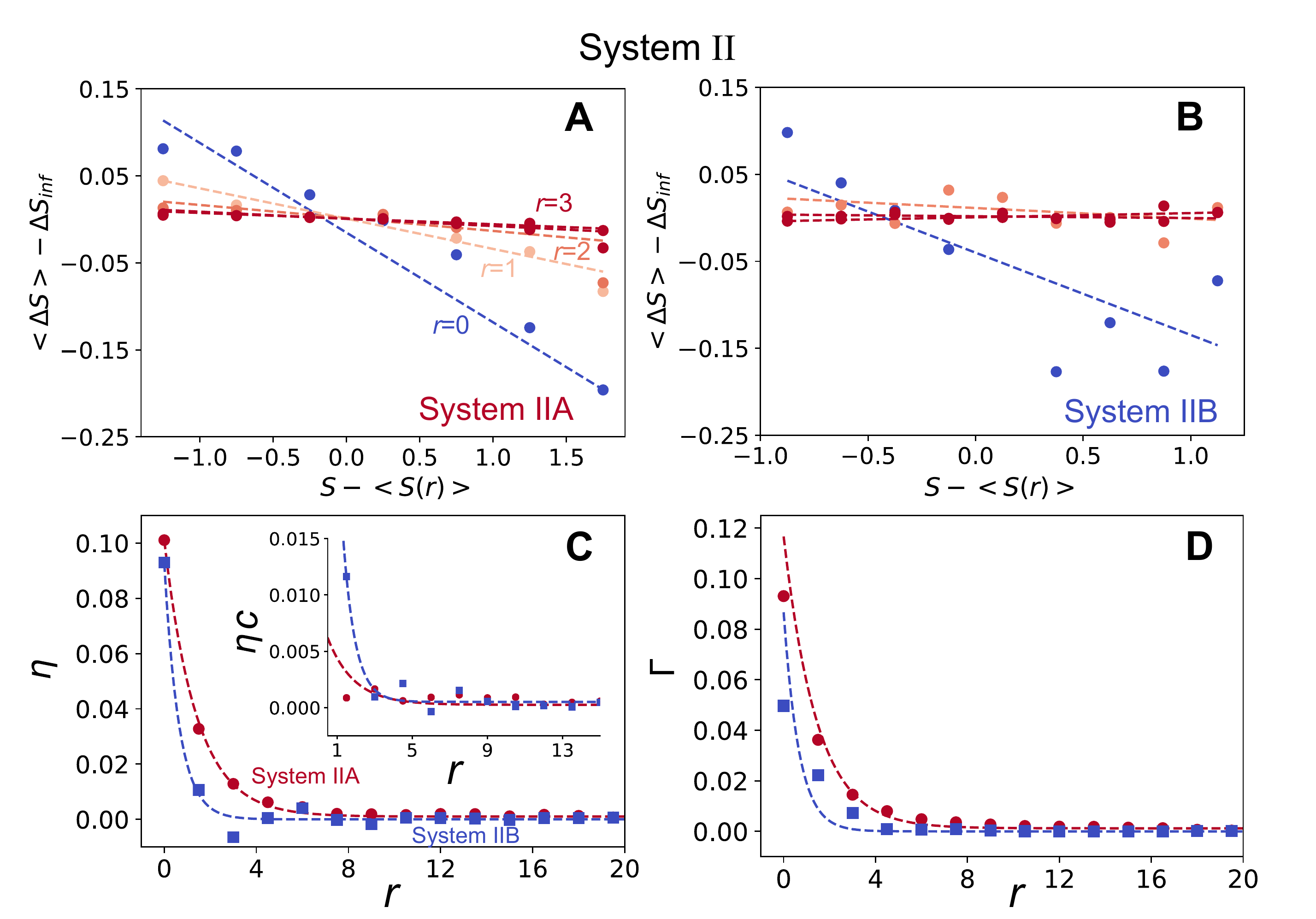}
\caption{Near-field contribution to softness change due to rearrangements in granular raft experiments (System II). (A) restoring effect for 1.0\,mm rafts (System IIA). $\Delta S_{\text{inf}}$ is the mean softness change in the background.  (B) restoring effect for 3.3\,mm rafts (System IIB). The dotted lines are linear fits. (C) restoring coefficient vs. distance to rearrangers. Inset: average softness change vs. distance. The curves are fits for System IIA, $\eta(r)=0.10e^{-r/1.40}$ and $c=0.08$, and for System IIB, $\eta(r)=0.09e^{-r/0.65}$ and $c=1.20$. (D) variance of softness change vs. distance to rearrangers.}
\label{fig:restore_granular}
\end{figure}

\begin{figure}[t]
\centering
\includegraphics[width=0.8\linewidth]{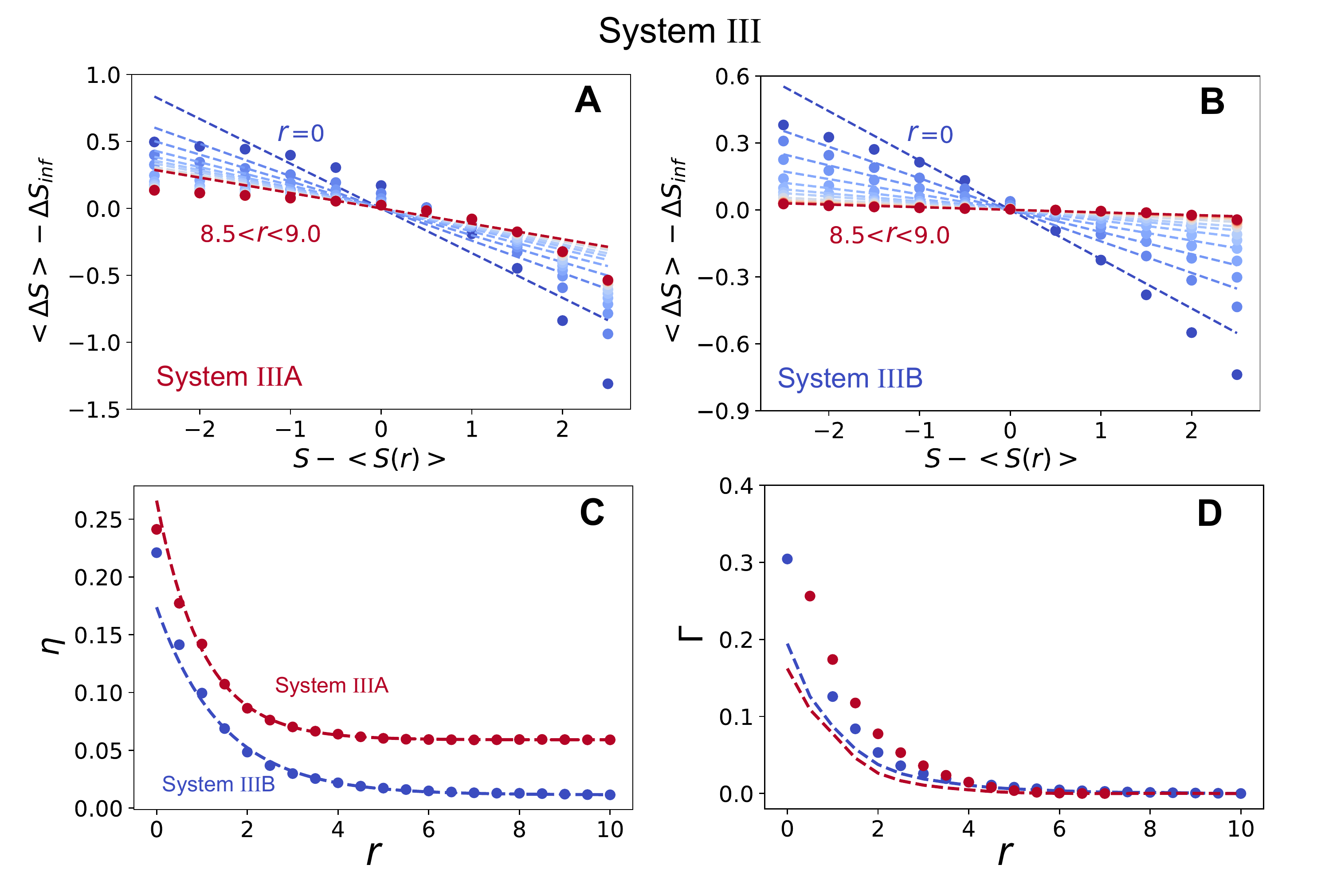}

\caption{Near-field structural contribution to softness change due to rearrangements in polymer nanopillars (System III). (A) restoring effect for $T=0.30$ (System IIIA). (B) restoring effect for $T=0.05$ (System IIIB).  (C) Restoring coefficient. The curves are fits for System IIIA, $\eta(r)=0.206e^{-r/1.02} + 0.059$, and for System IIIB, $\eta(r)=0.162e^{-r/1.45} + 0.011$. Notes that the $\eta$ decay length, $\xi_\eta$, scales with the $D_{min}^2$ correlation length, $\xi$, where $\xi_\eta / \xi \approx 1.33$ for both temperature. (D) Variance of softness change. Dot points are the variance measured in MD simulation and dash lines are the variance estimated from $\eta$.}
\label{fig:restore_oligomer}
\end{figure}

To measure the near-field softness kernel due to structural change induced by rearrangements, $\Delta S_{\text{struct}}=\eta(r)(\langle S \rangle  - S + c) + \delta(r)$, the change of the softness field $\Delta S(r) - \Delta S_\infty$ is linearly fitted as a function of the softness $S$ before the rearrangement~(37). This is evaluated at different distances to the rearranger and the results for the three respective systems are shown in Fig.~\ref{fig:restore_softDisk}A and B,~\ref{fig:restore_granular}A and B, and ~\ref{fig:restore_oligomer}A and B. For all three system, a negative slope $-\eta(r)$ is seen and $\eta$ decreases with the distance to the rearranger. For System I, we found a power-law decay fits $\eta(r)$ better, shown in Figure ~\ref{fig:restore_softDisk}C.
For Systems II and III, $\eta(r)$ is fitted to an exponential relation as shown in Fig.~\ref{fig:restore_granular}c and ~\ref{fig:restore_oligomer}C. In addition, a non-zero intercept $\eta(r)c$ exists for the granular experiments and the soft disk simulations, which is plotted in the inset of Fig.~\ref{fig:restore_softDisk}C and~\ref{fig:restore_granular}C.

The variance of the noise term, $\Gamma(r)$, follows a previously derived detailed balance-like relation, $\Gamma(r)=\eta(r)[2-\eta(r)]\sigma^2$, which describes the measured data well, see Fig.~\ref{fig:restore_softDisk}D, ~\ref{fig:restore_granular}D, and~\ref{fig:restore_oligomer}D.

\subsection{Far-field elastic contribution to softness kernel}

To calculate the parameters in the elastic contribution to the softness kernel, $\Delta S_{\text{elastic}}=\zeta\tilde{\epsilon}_{\text{vol}} + \kappa \Delta\left | \tilde \epsilon \right |^2$, a method based on global mean softness change was used~(37). 
Ensemble-averaged global mean softness, $\bar{S}(\epsilon_G)$, was calculated at different global strain. An assumption is made that at the very beginning of the deformation, all particles experience similar elastic strain and the plastic contribution of the few individual rearrangers is negligible. Under this assumption, we calculate $\zeta$ and $\kappa$ by fitting $\bar{S}(\epsilon_G)$ with $\zeta\epsilon_{\text{vol,G}} + \kappa \Delta\left | \epsilon_{\text{dev,G}} \right |^2$. Here, $\epsilon_{\text{vol,G}}=(1-\nu)\epsilon_G$ is the global volumetric strain and $\epsilon_{\text{dev,G}}=\frac{1+\nu}{2}\epsilon_G$ is the global deviatoric strain, and $\nu$ is the Poisson's ratio. For the 1.0\,mm granular raft (System IIA), $\nu=0.5$, $\zeta=2.37$, $\kappa=139$. For the 3.3\,mm granular raft (System IIB), $\nu=0.5$, $\zeta=1.74$, $\kappa=339$. 
Similar trend has been found in the polymer nanopillars (System III). For $T=0.30$ (System IIIA), $\zeta=9.17$, $\kappa=13.50$. For $T=0.05$ (System IIIB), $\zeta=16.60$, $\kappa=11.05$. 

For 2D repulsive disks (System I), since we applied a simple shear strain rather than a tensile strain, we could only find $\kappa$ from this fit. We found $\kappa=411$ fits both temperatures well. To extract $\zeta$, we measured locally-fit volumetric and deviatoric strains of particles during energy-minimization simulations. We performed fit $\Delta S_{\text{elastic}}=\zeta\tilde{\epsilon}_{\text{vol, local}} + \kappa \Delta\left | \tilde \epsilon_{\text{local}} \right |^2$, and found $\zeta=226$ for $T_a=0.025$ and $\zeta=244$ for $T_a=0.2$.

\subsection{Triggering probability for rearrangements}

\begin{figure}[t]
\centering
\includegraphics[width=1.0\linewidth]{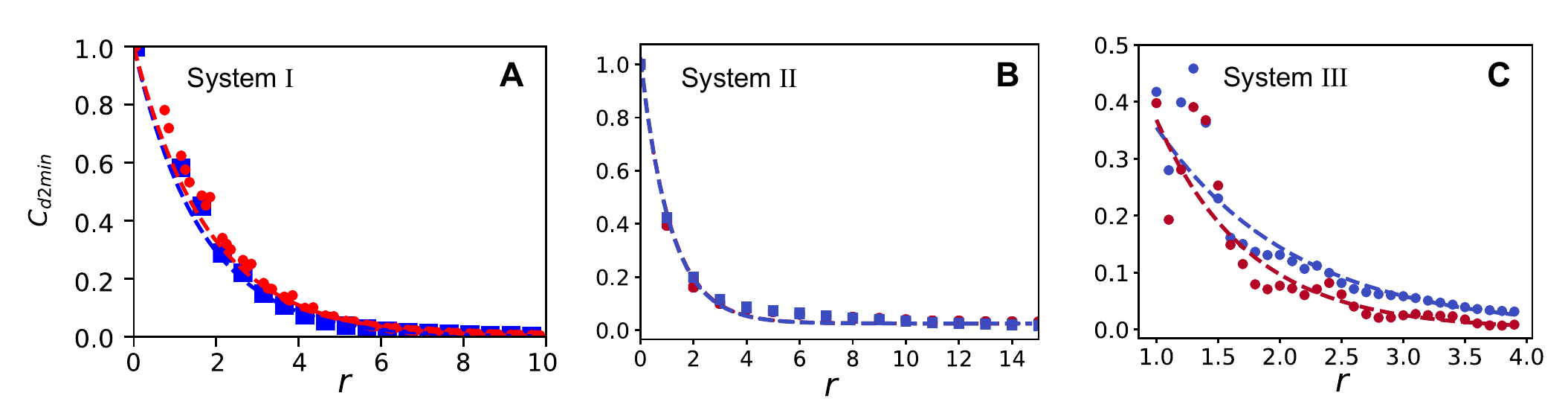}

\caption{Correlation of $D_{min}^2$ vs. distance to rearranger $r$. (A) Soft repulsive disks (System I), with $T_a=0.2$ (System IA) data fitted to $C_{d2min}(r)=1.0e^{-r/1.88}$ (red) and $T_a=0.025$ (System IB) data fitted to $C_{d2min}(r)=1.0e^{-r/1.63}$ (blue). (B) is same as (A), except for the granular rafts (System II), with 1.0\,mm data fitted to $C_{d2min}(r)=1.0e^{-r/1.13}$ (red), and 3.0\,mm data fitted to $C_{d2min}(r)=1.0e^{-r/1.12}$ (blue). (C) is same (A), except for the polymer nanopillars (System III), with $T=0.3$ (System IIIA) data fitted to $C_{d2min}(r)=1.40e^{-r/0.75}$ (red), and $T=0.05$ (System IIIB) data fitted to $C_{d2min}(r)=0.87e^{-r/1.11}$ (blue).}

\label{fig:d2mincorrelation}
\end{figure}

The rearrangements observed in the experiments and simulations typically involve a few particles. In the StEP simulation, each block has the length corresponding to the particle diameter. Thus, a plastic event in the StEP simulation should contain several blocks that rearrange at the same time. This is realized by letting the center rearranging block ($|\epsilon|>\epsilon_Y$) triggering nearby blocks to rearrange, with a probability $C(r)$. To approximate this probability, we use the spatial $D^2_{min}$ correlation at early stages of the deformation, $C(r)=C_{d2min}(r)$, and fit an exponential relation to the data, as shown in Fig.~\ref{fig:d2mincorrelation}. Note that for polymer nanopillars (System III), we do not force the correlation fitting to be continuous for the small range ($r < 1.0$), due to the existence of bonds. 

\begin{figure}[t]
\centering
\includegraphics[width=0.6\linewidth]{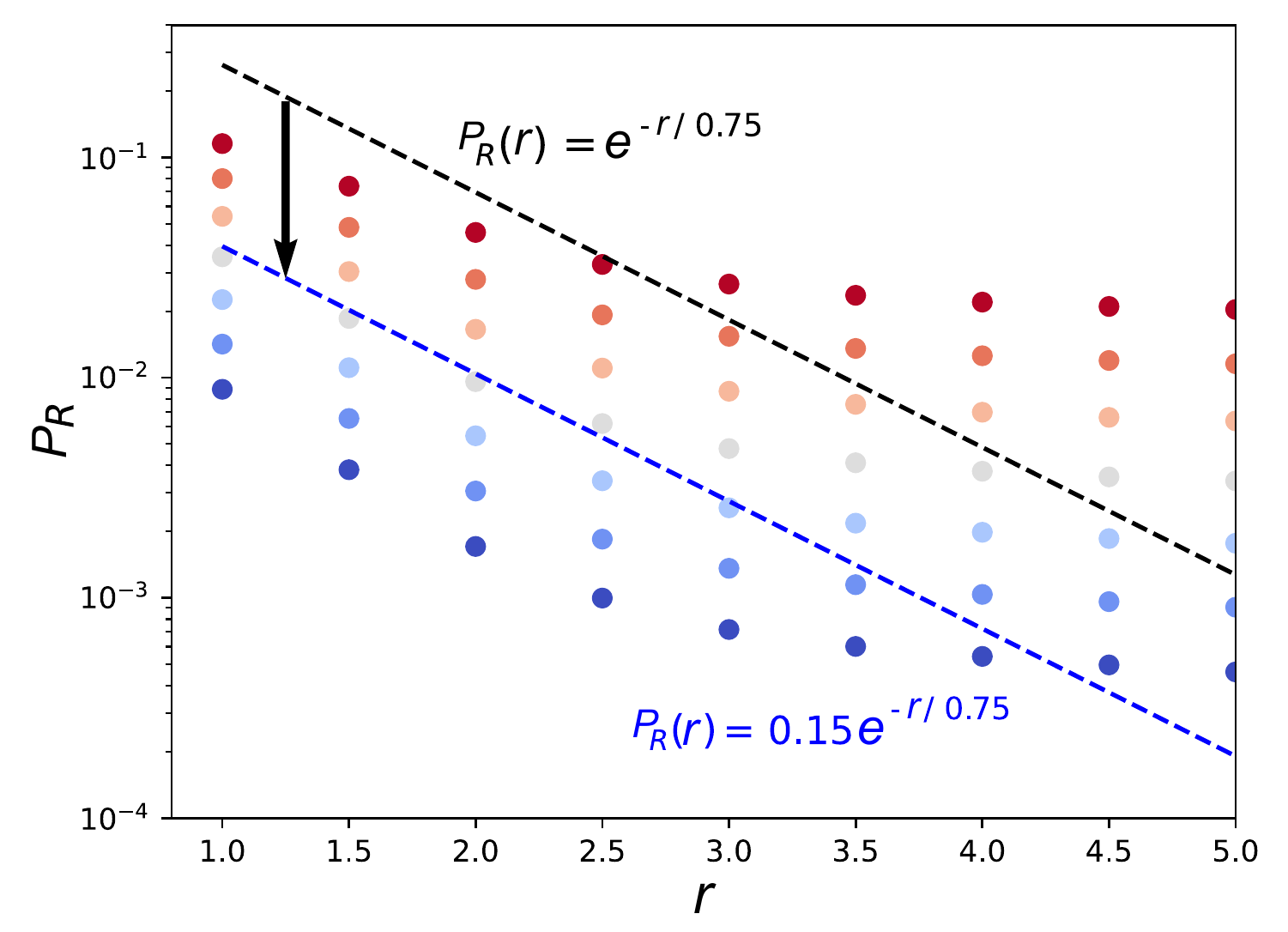}
\caption{Scaling $P_R$ at $T=0.30$ to make the rearranging probability in MD simulation and StEP model on the same order of magnitude. The color gradient represents the softness gradients from -3.0 (blue) to 3.0 (red). The dash lines are the calculated $P_R-P_{R, bulk}$ before and after scaling. We choose a scale factor of 0.15 and the same scale factor is then applied to pillars at $T=0.05$.}
\label{fig:thermal_scaled_d2min}
\end{figure}

For the thermal system, we also need to scale the term accounting rearrangement size when using the $D^2_{min}$ correlation length.
This is because $D^2_{min}$ measures the non-affine displacement and scales with temperature. 
In Figure \ref{fig:thermal_scaled_d2min}, we plotted the measured $P_R$ as a function of distance to the rearranging particle for different softness.
We can see that putting the correlation length directly overestimates $C(r)$, especially near the rearranging particle. 
Thus, we scale it to particles' rearranging probability, $P_R$, at the average softness, $P_R(\langle S \rangle)$, which gives us a scaling factor of $0.15$. The same scaling factor is then used for the low T as well.
We found that our results are qualitatively insensitive to the choice of this scaling factor, if the overall particle rearranging ratio in the StEP model is on the same order of magnitude as the corresponding MD simulation.

\begin{figure}[t]
\centering
\includegraphics[width=0.85\linewidth]{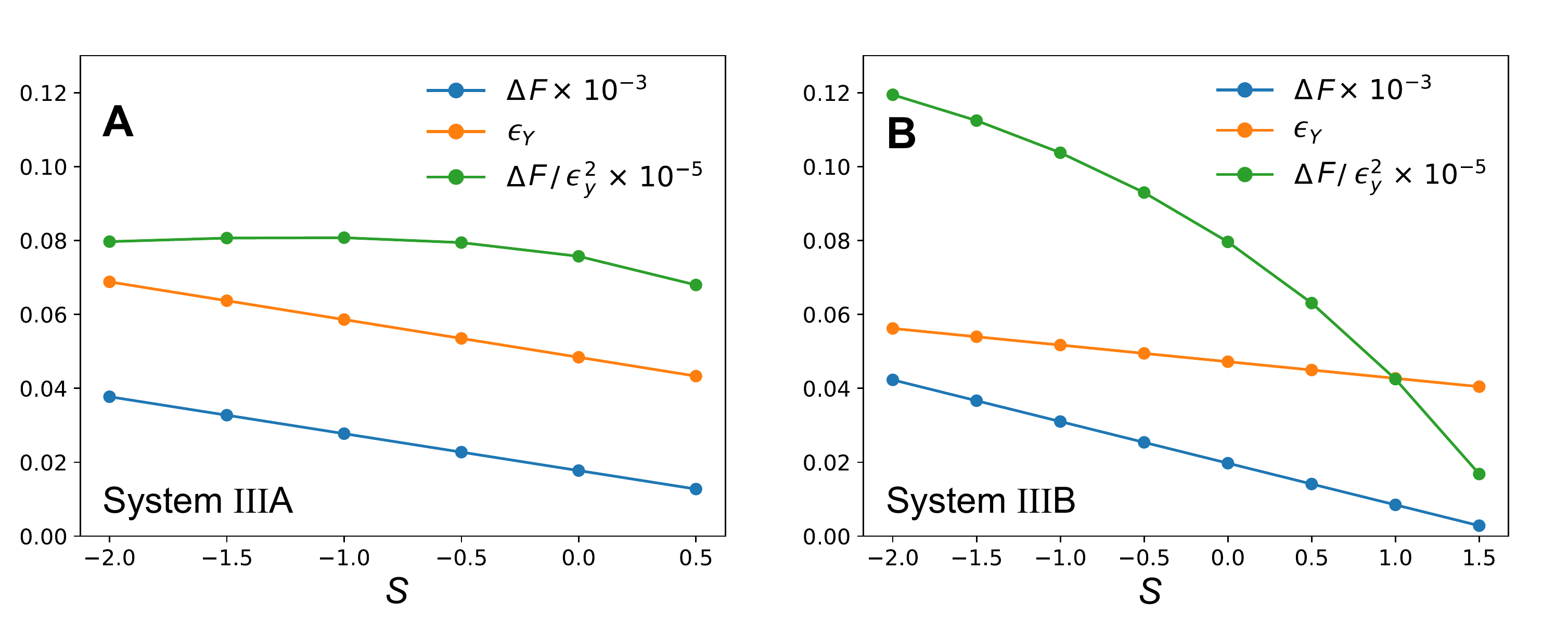}
\caption{\label{fig:thermal_eyring} Free energy barrier (blue), yield strain (orange), and their ratio (green) as a function of softness: (A) System IIIA with $T=0.30$; (B) System IIIB with $T=0.05$. At both temperature, $\Delta F (S) / \epsilon_y^2$ decreases with softness.}
\end{figure}

\subsection{Rearrangements triggered by thermal fluctuation}

The polymer nanopillars (System III) are different from the other two systems because they are at non-zero temperature.
To account for the thermal fluctuations that also induce rearrangements, a temperature-dependent rearranging mechanism is introduced for the polymer nanopillars. As explained before, a block starts rearranging if its elastic deviatoric strain, $\left | \tilde{\epsilon} \right |$, is larger than its local yield strain, $\epsilon_Y$. 
In the thermal system, if this rearranging criteria is not met, the site can also rearrange with a probability of $C_{th}$:
\begin{equation}
    C_{th} (S, \left | \tilde{\epsilon} \right |) = \exp(\Delta F (S) - V_0 G \left | \tilde{\epsilon} \right |^2) 
    \label{eq:thermal_re}  
\end{equation}
%

where $\Delta F$ is the free energy barrier for rearrangements in quiescent system, $V_0$ is the activation volume, $G$ is the local elastic modulus. 
The local elastic modulus is assumed to be the same for all the particles and equal to the  elastic modulus of the pillars, which are 29.13 in System IIIA and 74.58 in System IIIB. 
As shown by previous softness study~(28), $\Delta F (S)$ has the expression:
\begin{equation}
    \Delta F (S) = (\epsilon_0 - \epsilon_1 \cdot S) + T \cdot(- (\Sigma_0 - \Sigma_1 \cdot S)).
    \label{eq:thermal_F}  
\end{equation}
Here, $\epsilon_i$ and $\Sigma_i$ represent the enthalpic and entropic barrier for rearrangements, which are independent of both the temperature and the softness. 

Considering we have two rearranging mechanisms in the StEP model, we should expect that their predicted rearranging probabilities are continuous at $\left | \tilde{\epsilon} \right | = \epsilon_y$ for every particle. This gives out another constraint for the thermal yielding criteria:
\begin{equation}
    V_0 G = \Delta F (S) / \epsilon_Y^2.
    \label{eq:thermal_constiant}  
\end{equation}
We find that $\Delta F (S) / \epsilon_Y^2$ decreases with softness (see Fig.~\ref{fig:thermal_eyring}) at both temperatures. 
This trend meets our expectations because recent work shows that the activation volume, $V_0$, should decrease with softness~(50).